\documentclass[twocolumn,showpacs,preprintnumbers,amsmath,amssymb]{revtex4}

\usepackage{graphicx}
\usepackage{dcolumn}
\usepackage{bm}

\begin{document}

\preprint{}

\title{Anomalous in-plane magneto-optical anisotropy of self-assembled quantum dots}

\author{T.~Kiessling$^{1}$}
\author{A.~V.~Platonov$^{2}$}
\author{G.~V.~Astakhov$^{1,2}$}\email{astakhov@physik.uni-wuerzburg.de}
\author{T.~Slobodskyy$^{1}$}
\author{S.~Mahapatra$^{1}$}
\author{W.~Ossau$^{1}$}
\author{G.~Schmidt$^{1}$}
\author{K.~Brunner$^{1}$}
\author{L.~W.~Molenkamp$^{1}$}

\affiliation{ $^{1}$Physikalisches Institut (EP3) der
Universit\"{a}t
W\"{u}rzburg, 97074 W\"{u}rzburg, Germany \\
$^{2}$A.F.Ioffe Physico-Technical Institute, Russian Academy of
Sciences, 194021 St.Petersburg, Russia }

\date{\today}

\begin{abstract}
We report on a complex nontrivial behavior of the optical anisotropy
of quantum dots that is induced by a magnetic field in the plane of
the sample. We find that the optical axis either rotates in the
opposite direction to that of the magnetic field or remains fixed to
a given crystalline direction. A theoretical analysis based on the
exciton pseudospin Hamiltonian unambiguously demonstrates that these
effects are induced by isotropic and anisotropic contributions to
the heavy-hole Zeeman term, respectively. The latter is shown to be
compensated by a built-in uniaxial anisotropy in a magnetic field
$B_C = 0.4$~T, resulting in an optical response typical for
symmetric quantum dots.
\end{abstract}


\maketitle

Self assembled semiconductor quantum dots (QDs) attract much
fundamental and practical research interest. E.g., QD optical
properties are exploited in low-threshold lasers, while QDs are
also proposed as optically controlled qubits
\cite{qubit,Optic_qubit}. An important, and frequently somewhat
neglected aspect of QDs is the relationship between their symmetry
and their optical properties. We recently showed that extreme
anisotropy of QDs can lead to efficient optical polarization
conversion \cite{Conversion}.

Self assembled QDs grown by molecular beam epitaxy generally
possess a well-defined ($z$) axis along the $\mathrm{[001]}$
growth direction which serves as the spin quantization axis. It
thus makes sense to assume the dots have $D_{2d}$ point-group
symmetry, and to describe of the influence of an external magnetic
field using an isotropic transverse heavy-hole g-factor
$g_{hh}^{\perp}$ in the plane of the sample ($xy$). Many actual
QDs, however, exhibit an elongated shape in the plane and a
similar strain profile, their symmetry is reduced to $C_{2v}$ or
below. In this case the in-plane heavy hole g-factor is no longer
isotropic \cite{ani_g}. Moreover, even in zero field the
degeneracy of the radiative doublet is lifted due to the
anisotropic exchange splitting \cite{AlongatedShape, AniSplit,
Book}. Any of these issues will give rise to optical anisotropy,
resulting in the linear polarization of the photoluminescence
(PL).

In this manuscript, we discuss the optical anisotropy of QDs in
the presence of a magnetic field. Classically, the polarization
axis of the luminescence of a given sample should be collinear
with the direction of the magnetic field. This so-called Voigt
effect, implies that when rotating  the sample over an angle
$\alpha$ while keeping the direction of the magnetic field fixed,
one observes a constant polarization (in direction and amplitude)
for the luminescence. Mathematically one can express this behavior
as following a zeroth order spherical harmonic dependence on
$\alpha$. This situation changes drastically for low dimensional
heterostructures because of the complicated valence band
structure. Kusrayev \textit{et al.} \cite{Ku_Ku} observed a second
spherical harmonic component (i.e., $\pi$-periodic oscillations
under sample rotation) in the polarization of emission from narrow
quantum wells (QWs). This result was explained in terms of a large
in-plane anisotropy of the heavy-hole g-factor
$g_{hh}^{xx}=-g_{hh}^{yy}$ . Subsequently, this interpretation was
substantiated using a microscopic theory \cite{SemRya,QD_dich}.

Here, we report the observation of a fourth harmonic in the
magneto-optical anisotropy  (i.e. $\pi/2$-periodic oscillations in
the polarization of the emitted light under sample rotation) from
CdSe/ZnSe self assembled QDs. We demonstrate that this effect is
quite general. 
An important consequence is that the polarization axis hardly
follows the magnetic field direction, thus the classical Voigt
effect is not observed for QD emission associated with the
heavy-hole exciton. Moreover, in contrast to earlier studies of
in-plane magneto-optical anisotropies, we consider the
contributions of the electron-hole exchange interaction, which
have been ignored for QWs \cite{Ku_Ku,SemRya} and are zero for
charged QDs
\cite{ani_g}. 
An anisotropic exchange splitting may lead to the occurrence of a
compensating magnetic field $B_C$. When the externally applied field
equals $B_C$, the amplitude of the second harmonic crosses zero,
resulting in a highly symmetrical optical response of extremely
anisotropic QDs. For our QDs we find $B_C \approx 0.4$~T.
Intriguingly, at this condition the polarization axis rotates away
from the magnetic field direction.

\begin{figure}[tbp]
\includegraphics[width=.37\textwidth]{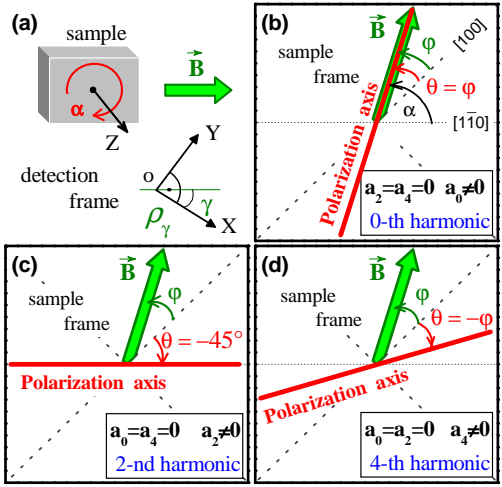}
\caption{(Color online) (a) Schematic layout of the angle-resolved
experiments where $\alpha$ is the rotation angle of the sample.
The detection frame $\rho_{\gamma}$ is rotated by an angle
$\gamma$ (equal to $0^{\circ}$ or $45^{\circ}$) with respect to
the magnetic field $\mathbf{B}$. (b)-(d) Different scenarios for
magneto-optical anisotropy (shown in the sample frame). (b) The
polarization axis follows the magnetic field and $\theta =
\varphi$, leading to a zeroth harmonic in the angle scan. (c) The
polarization axis is fixed at $\theta = \mathrm{const}(\varphi)$,
resulting in a second harmonic in the angle scan. (d) The
polarization axis rotates away from the magnetic field with
$\theta = -\varphi$, leading to a fourth harmonic in the angle
scan.} \label{fig1}
\end{figure}

In order to measure anisotropy we used the detection scheme
presented in Fig.~\ref{fig1}a. The direction of the in-plane
magnetic field is fixed, while the sample is rotated over an angle
$\alpha$. The degree of linear polarization is now defined as
$\rho_{\gamma}(\alpha) = (I_{\gamma} -
I_{\gamma+90^{\circ}})/(I_{\gamma} + I_{\gamma+90^{\circ}})$. Here
the angle $\gamma$ corresponds to the orientation of the detection
frame with respect to the magnetic field and $I_{\gamma}$ is the
intensity of the PL polarized along the direction $\gamma$. Within
an approximation of weak magnetic fields, a sample that has
$C_{2v}$ point symmetry in general may have only three spherical
harmonic components \cite{Ku_Ku} linking the polarization of the
emission to the sample rotation angle $\alpha$. Thus, one has for
$\gamma = 0^{\circ}$ and $\gamma = 45^{\circ}$
\begin{eqnarray}
\rho_{0} (\alpha) = a_0 + a_2 \, \cos  2\alpha + a_4 \, \cos
4\alpha \nonumber \\
\rho_{45} (\alpha) = a_2 \, \sin  2\alpha + a_4 \, \sin 4\alpha
\,, \label{eq1}
\end{eqnarray}
with $a_0$, $a_2$ and $a_4$ the amplitudes of the zeroth, second
and fourth harmonic, respectively.  $\alpha $ is measured with
respect to the $\mathrm{[1\overline{1}0]}$ crystalline axis.

For an analysis of the various orientation effects it is more
convenient to turn from the detection frame to the sample frame.
Now the sample orientation is fixed and the magnetic field rotates
by the same angle $\alpha$ but in opposite direction
(Figs.~\ref{fig1}b-d). The orientation of the magnetic field in
the spin Hamiltonian, which we discuss below, can be incorporated
more conveniently using a basis along the $\mathrm{[100]}$,
$\mathrm{[010]}$ and $\mathrm{[001]}$ crystalline axes. Therefore
we introduce the angle $\varphi$ between the magnetic field
direction and the $\mathrm{[100]}$ axis, as $\varphi \equiv \alpha
- 45^{\circ}$. We then consider the orientation of the
polarization axis described by an angle $\theta$ in the same
$\mathrm{[100]}$ basis.

We are now in a position to discuss some limiting cases of
Eq.~(\ref{eq1}). When the zeroth order spherical harmonic
dominates, $|a_0| \gg |a_2|, |a_4|$, the polarization axis of the
emission coincides with the magnetic field direction for any
orientation of the sample, $\theta (\varphi) =\varphi$
(Fig.~\ref{fig1}b). For a dominantly second harmonic response,
$|a_2| \gg |a_0|, |a_4|$, the polarization axis is fixed to a
distinct sample direction and does
not depend on the magnetic field orientation. 
In our experiments, the fixed polarization axis is
$\mathrm{[1\overline{1}0]}$ and $\theta (\varphi) = -45^{\circ}$
(Fig.~\ref{fig1}c). The most interesting behavior is observed for
the fourth harmonic $|a_4| \gg |a_0|, |a_2|$, when the
polarization $\rho_{\gamma}(\alpha)$ in the detection frame
changes twice as fast as any polarization linked to the sample
frame. This implies that the polarization axis turns in opposite
direction to that of the magnetic field, and $\theta (\varphi)
=-\varphi$ (Fig.~\ref{fig1}d). In other words, it is collinear
with the magnetic field when $\mathbf{B} \parallel \mathrm{[100]},
\mathrm{[010]}$ and perpendicular to the magnetic field when
$\mathbf{B}
\parallel \mathrm{[110]}, \mathrm{[1\overline{1}0]}$.

\begin{figure}[tbp]
\includegraphics[width=.41\textwidth]{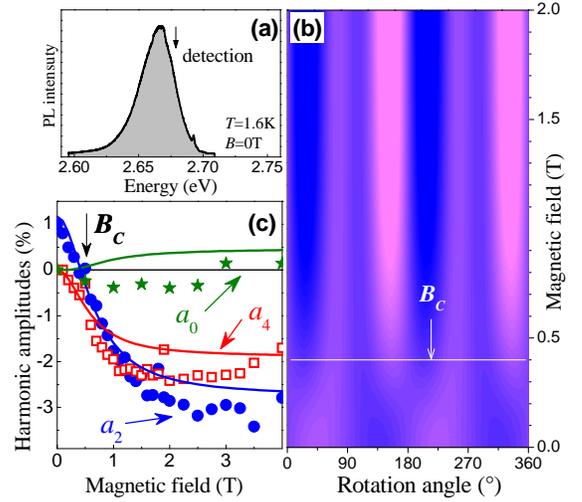}
\caption{(Color online) (a) PL  emission spectrum of the CdSe/ZnSe
QD sample. 
(b) 3D plot of the linear polarization $\rho_{45}(\alpha,B)$ as
function of the rotation angle $\alpha$ and magnetic inductance
$B$. Light (red) and dark (blue) areas correspond to positive and
negative values of $\rho_{45}(\alpha,B)$, respectively. (c)
Amplitudes of the zeroth ($a_0$), second ($a_2$) and fourth
($a_4$) spherical harmonics \textit{vs} $B$. The symbols are
experimental data and the solid lines result from calculations.
The arrows in panels (b) and (c) indicate the compensating field
$B_C$ where the linear polarization is $\pi/2$-periodic ($a_2 =
0$). } \label{fig2}
\end{figure}

We studied the magneto-optical anisotropy of CdSe QDs in a ZnSe
host. The samples were fabricated on (001) GaAs substrates using
molecular beam epitaxy and self-assembly after depositing one
monolayer of CdSe on a 50 nm thick ZnSe layer. The QDs were then
capped by 25~nm of ZnSe. For optical excitation at 2.76~eV we used
a stilbene dye-laser pumped by the UV-lines of an Ar-ion laser.
 A typical PL spectrum 
is shown in Fig.~\ref{fig2}a. For detection of the linear
polarization we applied a standard technique using a photo-elastic
modulator and a two-channel photon counter \cite{Conversion}. The
angle scans were performed with the samples mounted on a rotating
holder controlled by a stepping motor with an accuracy better than
$1^{\circ}$. Magnetic fields up to 4~T were applied in the sample
plane (Voigt geometry), optical excitation was done using a
depolarized laser beam. The degree of linear polarization of the
luminescence of the QDs  was determined at 2.68~eV at a
temperature $T=1.6$~K.

\begin{figure}[tbp]
\includegraphics[width=.39\textwidth]{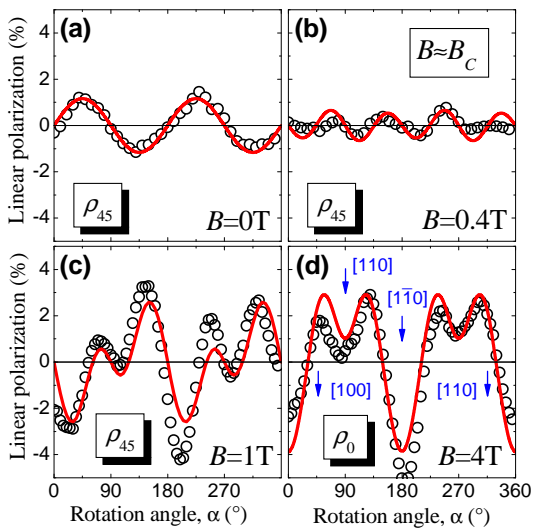}
\caption{(Color online) Angle scans of the linear polarization of
the luminescence. The symbols are experimental data, solid lines
represent calculations based on the Hamiltonian of
Eq.~(\ref{eq7Main}). (a) Built-in linear polarization ($B=0$~T). (b)
A highly symmetrical optical response (i.e. $\pi/2$-periodic)
appears at a compensating field $B_c \approx 0.4$~T.  (c) Angle scan
in a magnetic field $B=1$~T exeeding $B_C$. (d) Angle scan in the
saturation regime, $B=4$~T. The detection frame $\rho_{\gamma}$ in
(a)-(c) is rotated by an angle $\gamma = 45^{\circ}$; in (d) $\gamma
= 0^{\circ}$. The arrows indicate the points where the orientation
of the magnetic field aligns with high-symmetry directions in the
sample frame.} \label{fig3}
\end{figure}

The result of angle scans of $\rho_{45}$ for various magnetic field
strengths is shown in Fig.~\ref{fig2}b. In zero magnetic field the
linear polarization is $\pi$-periodic (see also Fig.~\ref{fig3}a).
This demonstrates the low ($C_{2v}$) symmetry of the QDs (the dots
are elongated in the plane) resulting in a finite value of $a_2$ in
Eq.~(\ref{eq1}). One can regard this measurement as reflecting the
'built-in' linear polarization of the array of dots. For high
magnetic fields $a_2$ changes sign and an additional fourth harmonic
signal appears (see also Fig.~\ref{fig3}c). At an intermediate
magnetic field $B_C \approx 0.4$~T the second harmonic crosses zero,
while the fourth harmonic remains finite (see also
Fig.~\ref{fig3}b). The resulting $\pi/2$-periodic optical response
corresponds to the higher ($D_{2d}$) symmetry expected for symmetric
(i.e., not elongated) QDs. This observation clearly demonstrates
that the in-plane optical anisotropy of QDs can be compensated by
in-plane Zeeman terms. We also find that the zeroth harmonic signal
is weak. This is clearly seen e.g. for the $\rho_{0}$ angle scan in
Fig.~\ref{fig3}d  which was taken
for 
$B=4$~T, where the field-induced alignment is saturated.

We now model our observations using a pseudospin formalism
\cite{SemRya,LinPolDMS}. We denote by $|+Sp\rangle$ and
$|-Sp\rangle$ the wavefunctions of an electron ($p=el$) or
heavy-hole ($p=hh$) with pseudospin projection $\pm S$ along the
$z$ direction. We define a pseudospin Hamiltonian in matrix form
for each band, as follows
\begin{equation}
\mathcal{H}_p = \frac{\delta_{p}}{2} (\sigma_{x} \cos \theta_p +
\sigma_{y} \sin \theta_p ) \,,\label{eq2}
\end{equation}
where $\sigma _x$ and $\sigma _y$ are the Pauli matrices.
This Hamiltonian has eigenfunctions
$\Psi_p^{\pm 1} \propto |+Sp\rangle \pm e^{i\theta_p}
|-Sp\rangle$. The optical matrix elements for a transition between
electron and heavy-hole bands are $\langle \pm\frac{1}{2}el |
\hat{P} | \pm\frac{3}{2}hh \rangle = \mp e_{\pm}$ \cite{Book},
where $\hat{P}$ is the dipole momentum operator. We define
$e_{\pm}=(e_{x} \pm i e_{y})/\sqrt{2}$ with $e_{x} \parallel
[100]$ and $e_{y} \parallel [010]$ are unit vectors. We thus find
for the optical matrix elements for the four possible optical
transitions 
\begin{equation}
\langle \Psi_{el}^{\eta} | \hat{P} | \Psi_{hh}^{\mu} \rangle
\propto - e_{+} + \eta \mu \, e^{i(\theta_{hh} - \theta_{el}) }
e_{-} \,.\label{eq3}
\end{equation}
These matrix elements thus predict a linear polarization with an
axis that is rotated over an angle $\theta =
\frac{1}{2}(\theta_{hh} - \theta_{el})$ from [100] when
$\eta=-\mu$ ($\eta, \mu = \pm 1$), and rotated over $\theta =
\frac{1}{2}(\theta_{hh} - \theta_{el})+ 90^{\circ}$ when
$\eta=\mu$.

For an electron in an external magnetic field $B$, we can write
the Zeeman Hamiltonian as
\begin{equation}
\mathcal{H}_{el} = \frac{1}{2} g_{el}^{\bot} \mu_B  \left(
\sigma_x B \cos \varphi + \sigma_y B \sin \varphi \right) \,,
\label{eq4}
\end{equation}
where $g_{el}^{\bot}$ is the electron g-factor, yielding directly
$\theta_{el} = \varphi$. For $D_{2d}$ symmetry, the interaction of
holes with an in-plane magnetic field $B$ can be described by
\cite{HoleSplit}
\begin{equation}
\mathcal{H}_{hh} =  q_1 \, g_0 \mu_B  \left( J_x^3 B \cos \varphi
+ J_y^3 B \sin \varphi \right)  \,, \label{eq5}
\end{equation}
where $q_1$ is a constant. It suffices to note that the part of
matrices $J_x^3$ and $J_y^3$, related to the angular momentum of
heavy holes, behave as $\frac{3}{4}\sigma_x$ and
$-\frac{3}{4}\sigma_y$, respectively \cite{Book}.
On comparing the resulting Hamiltonian with Eq.~(\ref{eq2}) this
implies $\theta_{hh} = - \varphi$ for the eigenstate, yielding
[cf. Eq.~(\ref{eq3}) and the text thereafter] a rotation over an
angle $\theta = -\varphi$ or $\theta = -\varphi + 90^{\circ}$ for
the polarization of the luminescence. In accordance with
Fig.~\ref{fig1}d this corresponds to the fourth spherical
harmonic.

The second harmonic may appear for structures with $C_{2v}$ symmetry
or below. In this case there is a correction to the magnetic
interaction, given by \cite{HoleSplit}
\begin{equation}
\mathcal{H}_{hh}' =  q_2 \, g_0 \mu_B  \left( J_x^3 B \sin \varphi
+ J_y^3 B \cos \varphi \right) \,, \label{eq6}
\end{equation}
where $q_2$ is $C_{2v}$ invariant. From this corection one finds
$\theta_{hh} = \varphi - 90^{\circ}$, leading to a rotation of the
luminescence polarization over $\theta = \pm 45^{\circ}$, which
implies (Fig.~\ref{fig1}c) a response following a second spherical
harmonic.

For a quantitative analysis a more detailed approach is required
which we provide below. The essential experimental data are
summarized in Fig.~\ref{fig2}c, where we plot the amplitudes of
the spherical harmonic, i.e. the coefficients $a_0$, $a_2$ and
$a_4$, extracted from the fits of the experimental data on
$\rho_\gamma(\alpha)$ using Eq.~(\ref{eq1}) \textit{vs} magnetic
field (symbols).

In QDs, the electron-hole exchange interaction is significant and
a corresponding term $\mathcal{H}_{ex}$ must be taken into account
\cite{QD_Voigt}, resulting in an exchange splitting $\delta_0$
between the $| L_z = \pm 1 \rangle$ and $| L_z = \pm 2 \rangle$
exciton states, where $L_z$ is the projection of the angular
momentum of the exciton. Moreover, $\mathcal{H}_{ex}$ leads to a
(smaller) splitting $\delta_2 < \delta_0$ of the nonradiative
exciton states. When the symmetry is lowered to $C_{2v}$, an
additional anisotropic exchange term $\mathcal{H}_{ex}'$
\cite{SpinHam,Book} appears and results in an additional splitting
$\delta_1$ of the radiative doublet.

In the basis of the exciton states $\Phi_{1,2}=| L_z = \pm 1
\rangle$ and $\Phi_{3,4}=| L_z = \pm 2 \rangle$, the final spin
Hamiltonian $\mathcal{H} = \mathcal{H}_{el} + \mathcal{H}_{hh} +
\mathcal{H}_{hh}' + \mathcal{H}_{ex} + \mathcal{H}_{ex}'$ is given
by the following matrix \cite{Book,SpinHam}
\begin{equation}
\mathcal{H} = \frac{1}{2}  \left(%
\begin{array}{cccc}
  \delta_0 & -i \delta_1 & \delta_{el} & \delta_{hh} \\
  i \delta_1 & \delta_0 & \delta_{hh}^{*} & \delta_{el}^{*} \\
  \delta_{el}^{*} & \delta_{hh} & -\delta_0 & \delta_2 \\
  \delta_{hh}^{*} & \delta_{el} & \delta_2 & -\delta_0 \\
\end{array}%
\right) \,. \label{eq7Main}
\end{equation}
Here, $\delta_{el} = \mu_B g_{el}^{\bot} B_{+}$ and $\delta_{hh} =
\mu_B \left( g_{hh}^{\mathrm{i}} B_{+} + i g_{hh}^{\mathrm{a}}
B_{-} \right)$ are in-plane Zeeman terms,  $g_{hh}^{\mathrm{i}} =
\frac{3}{2} g_0 q_1$ and $g_{hh}^{\mathrm{a}} = \frac{3}{2} g_0
q_2$ are the isotropic and anisotropic contributions to the
heavy-hole g-factor. $B_{\pm} = Be^{\pm i \varphi}$ are effective
magnetic fields. Analytical solutions for the energy eigenvalues
$E_j$ and the normalized eigenfunctions $\Psi_j=\Sigma
V_{nj}\Phi_n$  of Hamiltonian~(\ref{eq7Main}) can be found in the
high magnetic field limit ($|\delta_{hh}|, |\delta_{el}| \gg
|\delta_0|, |\delta_1|, |\delta_2|$), as follows
\begin{equation}
E_{1,4}=\mp|\delta_{el}| \mp |\delta_{hh}| \,\mathrm{;} \,\,\,\,\,
E_{2,3}=\pm|\delta_{el}| \mp |\delta_{hh}| \,, \nonumber
\end{equation}
\begin{equation}
V=\frac{1}{2}\left(
\begin{array}{cccc}
1 & 1 & 1 & 1 \\
e^{2i\theta} & -e^{2i\theta} & - e^{2i\theta} & e^{2i\theta}  \\
-e^{-i \theta_{el}} & e^{-i \theta_{el}} & - e^{-i \theta_{el}} & e^{-i \theta_{el}} \\
-e^{i \theta_{hh}} &  -e^{i \theta_{hh}} & e^{i \theta_{hh}} & e^{i \theta_{hh}} \\
\end{array}
\right), \label{eq8Anal}
\end{equation}
where $e ^{i \theta_{el}} = \delta_{el} / |\delta_{el}|$, $e ^{i
\theta_{hh}} = \delta_{hh}^* / |\delta_{hh}|$ and $e^{2i\theta}=e
^{i (\theta_{hh}-\theta_{el})}$ have the same meaning as in
Eq.~(\ref{eq3}). Eqs.~(\ref{eq8Anal}) clearly show the same trends
expected from the qualitative analysis presented above
[Eqs.~(\ref{eq2}-\ref{eq6})]. For a given finite magnetic field,
solutions for $E_j$ and $\Psi_j$ can easily be calculated
numerically.

Using the solutions to Hamiltonian~(\ref{eq7Main}) we may obtain the
intensity and polarization of the luminescence. Since only the
$\Phi_{1,2} = | L_z = \pm 1 \rangle$ excitons are optically active,
the optical matrix element in an arbitrary direction $\mathbf{e}$
for eigenfunction $\Psi_j$ can be written as: $M_j (\mathbf{e})
=-V_{1j}e_{+} + V_{2j}e_{-}$ \cite{SpinHam}. We then can calculate
the intensity of the luminescence linearly polarized along an axis
rotated by an angle $\xi$ with respect to the [100] crystalline axis
as $I_{j,(\xi)}=|M_j(\mathbf{e}||{\xi})|^2$.

The polarization of the PL from an ensemble of QDs must be
averaged over the thermal population of exciton states, and can in
the sample frame be expressed as
\begin{equation}
\rho'_{(\xi)} = K \frac{\sum_j \, P_j \left( I_{j,(\xi)} -
I_{j,(\xi +90^{\circ})} \right)}{\sum_j \, P_j \left( I_{j,(\xi)}
+ I_{j,(\xi +90^{\circ})} \right)} \,, \label{eq10}
\end{equation}
where $P_j \propto e^{-E_j/k_B T}$ is the Boltzmann factor .
$K$ is a scaling factor which corrects for spin relaxation, and
basically determines the saturation level of the polarization at
high magnetic fields.

Using this approach, we have calculated the linear polarizations
$\rho'_{(100)}$ and $\rho'_{(1\overline{1}0)}$ in the sample frame
using Eq.~(\ref{eq10}). From this, we found the polarizations
$\rho_{0} (\alpha) = \rho'_{(1\overline{1}0)} \cos 2\alpha +
\rho'_{(100)} \sin 2\alpha $ and $\rho_{45} (\alpha) =
\rho'_{(1\overline{1}0)} \sin 2\alpha - \rho'_{(100)} \cos 2\alpha$.
We were able to reproduce all experimental data using a unique set
of parameters for given excitation power and excitation energy. The
calculations were done taking a bath temperature $T=1.6$~K and a
coefficient $K=0.04$. From the best fits we found exchange energies
$\delta_0 = 2.9$~meV, $\delta_2 = 0.1$~meV, $\delta_1 = 0.2$~meV and
g-factors $g_{e}^{\bot} = 1.1$, $g_{hh}^{\mathrm{i}} = -0.5$,
$g_{hh}^{\mathrm{a}} = 0.6$ .

An exemplary result of the calculations are the solid lines in
Fig.~\ref{fig3}, which follow  the experimental data very closely.
For a large number of similar fits we have evaluated the harmonic
amplitudes ($a_0$, $a_2$ and $a_4$) using Fourier transformation.
The results as a function of magnetic field are plotted as solid
lines Fig.~\ref{fig2}c. From this plot we find a compensating
field $B_C$ of 0.42~T, which coincides within the detection error
with the experimental value. We observe that as a general trend
the isotropic hole g-factor $g_{hh}^{\mathrm{i}}$ is responsible
for the fourth harmonic and the anisotropic g-factor
$g_{hh}^{\mathrm{a}}$ is essential for the second harmonic. The
latter was reported previously for magnetic QWs \cite{Ku_Ku} and
our findings follow the general trend of previous theoretical
approaches \cite{SemRya}. We conclude that the very good agreement
with experiment proves the validity of our approach.

Summarizing, we have observed anomalous behavior of the in-plane
magneto-optic anisotropy in CdSe/ZnSe QDs, in that the second and
fourth spherical harmonics of the response dominate over the
classical zeroth order response. We show the existence of a
compensating magnetic field, leading to a symmetry enhancement of QD
optical response. All of these findings could be excellently modeled
using a pseudospin Hamiltonian approach, and provide direct evidence
for the existence of a preferred crystalline axis for the optical
response, and for the importance of the exchange interaction for the
optical properties of self assembled QDs. The physics responsible
for our findings is not limited to QDs and can be applied to other
heterostructures of the same symmetry where the heavy-hole exciton
is the ground state.

This work was supported  by the Deutsche Forschungsgemeinschaft
(SFB 410) and RFBR.


\end{document}